\documentclass[]{article}

\usepackage[]{graphicx}
\usepackage{color}

\newcommand{\ket}[1]{\left| #1 \right\rangle}
\newcommand{\bra}[1]{\left\langle #1\right |}

\begin{document}
\setlength{\textheight}{8.0truein}  

\thispagestyle{empty}
\setcounter{page}{1}

\vspace*{0.88truein}

\centerline{\bf PROPOSAL FOR REALIZATION OF A TOFFOLI GATE}
\vspace*{0.035truein}
\centerline{\bf VIA CAVITY-ASSISTED ATOMIC COLLISION}

\vspace*{0.37truein}

\centerline{\footnotesize HAROLD OLLIVIER}

\vspace*{0.015truein}

\centerline{\footnotesize\it INRIA, Projet Codes, BP 105 Domaine de Voluceau,} 
\baselineskip=10pt
\centerline{\footnotesize\it F-78153 Le Chesnay cedex, France}
\baselineskip=10pt
\centerline{\footnotesize\it and}
\baselineskip=10pt
\centerline{\footnotesize\it Los Alamos National Laboratory, T-DO/QC, MS B213} 
\baselineskip=10pt
\centerline{\footnotesize\it Los Alamos, New Mexico 87545, USA}

\vspace*{10pt}

\centerline{\footnotesize PEROLA MILMAN}

\vspace*{0.015truein}

\centerline{\footnotesize\it Laboratoire Kastler Brossel, D\'epartement de Physique de l'Ecole Normale Sup\'erieure,}
\baselineskip=10pt
\centerline{\footnotesize\it 24 rue Lhomond, F-75231 Paris Cedex 05, France}
\baselineskip=10pt
\centerline{\footnotesize\it and}
\baselineskip=10pt
\centerline{\footnotesize\it Coll\`ege de France, 11 place Marcelin-Berthelot,}
\baselineskip=10pt
\centerline{\footnotesize\it F-75005 Paris, France}
\vspace*{0.225truein}

\vspace*{0.21truein}

\abstract{Cavity {QED} is a versatile tool to explore small
scale quantum information processing. Within this setting, we describe
a particular protocol for implementing a Toffoli gate with Rydberg
atoms and a cavity field. Our scheme uses both resonant and non
resonant interactions, and in particular a cavity assisted atomic
collision. The experimental feasibility of the protocol is carefully
analyzed with the help of numerical simulations and takes into account
the decoherence process. Moreover, we show that our protocol is
optimal within the constraints imposed by the experimental
setting.}

\vspace*{10pt}

\vspace*{3pt}

\section{Introduction}        

The recent development of quantum information processing has shed new
light on complexity and communication theory. It is now widely
accepted that some problems are solved more efficiently by quantum
computers than by their classical analogues~\cite{Sho94a,
Gro96a}. This matter of fact has triggered in the past years a lot of
studies on theoretical and practical aspects of quantum
computation. In particular, finding universal sets of gates for
processing quantum information is of paramount importance: the
possibility of implementing universal gates is a key requirement for
building interesting quantum information processing devices. A lot of
efforts have been dedicated to this question, leading to major
theoretical results (for a review see~\cite{NC00a}).

One of the next challenges that need to be overcome is to design
robust implementations of these sets of gates. Several constructions
have already been proposed and realized for various experimental
settings~\cite{THLM95a, RNOB99a, MMKI95a, GC97a, JMH98a}. These
results currently serve as benchmarks for other implementations and do
provide deep insights into the ability of the chosen devices to
manipulate quantum systems efficiently. In this article, we describe a
protocol for realizing the Toffoli gate --- a three qubits
``control-control-NOT'' --- in the Cavity {QED} context
({CQED}). This gate, together with one qubit rotations, form a
universal set for quantum computing~\cite{NC00a}. We will see
that the combination of techniques used makes our protocol optimal in
term of number of interactions within the constraints of our
experimental setting. It is thus less resources demanding than the
standard implementation as a sequence of control-NOT
gates~\cite{DS94a}. Finally we estimate the performance of our
protocol by taking into account imprecisions as well as decoherence
effects during the protocol.

\section{Cavity-{QED} toolbox}

In this paragraph we review briefly the different techniques used for
implementing the Toffoli gate. Further experimental details concerning
our particular setting are exposed with great care in
\cite{RBH01a}. In our protocol, quantum information will be stored in
circular Rydberg states of highly excited Rubidium atoms and in the
cavity field. The very long lifetime of these atoms together with the
transition frequencies of the chosen transitions allow to deal with
those complex atoms as if they had only three levels noted $\ket g$,
$\ket e$ and $\ket i$. Transformations between those three states can
be driven coherently by a classical microwave pulse adjusted nearly
resonantly to the proper transition frequency. For example, if the
classical field is tuned with respect to $\ket i \leftrightarrow \ket
g$, a $\pi$ pulse will transform any pure quantum state of the form
$\alpha \ket e + \beta \ket g$ into $\alpha \ket e + \beta \ket
i$. Indeed, those transformations are the analogues of the well known
one qubit gates for our three level systems. In order to process
quantum information in a non-trivial way, we also need an equivalent
for the two-qubit gates. Within the {CQED}
context, this involves a coupling between the atoms and a high-{\em Q}
Niobium superconducting microwave cavity. The frequency of the mode
inside the cavity can be adjusted in and out of resonance with the
$\ket g \leftrightarrow \ket e$ transition. Thus, we can consider two
distinct approaches for processing quantum information. The first one,
which has been throughly used in recent experiments~\cite{NROB99a},
relies on a resonant interaction with a single mode of the cavity. As
an example, for a well chosen interaction time, an initial atom-field
state of the form $(\alpha \ket g + \beta \ket e)\ket 0$ will evolve
into $\ket g (\alpha \ket 0 + \beta \ket 1)$. This transformation is
called $\pi$-Rabi rotation. Similarly, continuing the interaction for
an equal amount of time leads to the state $(\alpha \ket g - \beta
\ket e)\ket 0$. In the perspective of information processing it
corresponds to transferring the information from the atom into the
cavity and back to the atom. Hence, the field inside the cavity acts
as a temporary quantum memory. The second approach for building atom
field interactions, referred to as cavity assisted atomic collision,
has been proposed~\cite{ZG00a, Zhe01a} and experimentally
tested~\cite{OBAM01a} more recently. In this setting, two atoms enter
the cavity at the same time and follow an evolution conditioned upon
the state of the cavity field. More precisely, the cavity is detuned
from the $\ket e \leftrightarrow \ket g$ transition. In this regime,
where the detuning $\delta$ is much larger than the atom-field
coupling constant $\Omega$, the effective Hamiltonian can be derived
using second order perturbation theory:
\begin{eqnarray}
H_e & = & \lambda \left(\ket {e_1} \bra {e_1} aa^+ - \ket {g_1}\bra {g_1} a^+a \right. \nonumber \\
&& +\, \ket {e_2} \bra {_2}e aa^+ - \ket {g_2}\bra {g_2} a^+a \\ 
&& \left.+\, \ket {e_1} \bra {g_1} \otimes \ket {g_2} \bra {e_2} + \ket {g_1} \bra {e_1} \otimes \ket {e_2} \bra {g_2}\right), \nonumber \label{Heff}
\end{eqnarray}
where $\lambda = \Omega^2/4\delta$, and where the operators $a$ and
$a^+$ are the annihilation and creation operators for the cavity
field. In the above formula, we see that any exchange of energy
between the field and the atoms present in the cavity is forbidden,
but this still allows conditional dynamics upon the photon number in
the cavity mode. This interaction will be our main tool to perform the
Toffoli gate in the {CQED} setting.

\section{Description of the protocol}
Before going into the details of the protocol, recall that the Toffoli
gate is a three qubits gate. Its action on the computational basis of
the three qubits is to perform a ``control-control-NOT''. Thus, only
two basis vectors are affected by this evolution:
\begin{equation}
\begin{array}{rcl}
\ket 1 \ket 1 \ket 0 & \rightarrow \ket 1 \ket 1 \ket 1 \\
\ket 1 \ket 1 \ket 1 & \rightarrow \ket 1 \ket 1 \ket 0 
\end{array}
\end{equation}
where the control qubits are the first two ones, and the target is the
last one. Taking into account the specific {CQED} setting, we
chose an implementation with one cavity mode and two atoms. The photon
number states of the cavity mode will be denoted $\ket{n_c}$, while
the energy levels of the atoms will be written as $\ket {i_{c,t}},
\ket {g_{c,t}}, \ket{e_{c,t}}$ (the subscripts $c$ and $t$ are short
hands for control and target). The relation between those states and
the computational basis of the Toffoli gate is summarized below
(obvious normalization factors have been omitted):
\begin{equation}
  \begin{array}{c||c|c|c}
\textrm{Computational}&\textrm{Control 1} & \textrm{Control 2} & \textrm{Target} \\
\textrm{basis}        &\textrm{(cavity mode)} & \textrm{(Rb atom)} & \textrm{(Rb atom)} \\ \hline\hline
    \ket 0 & \ket {1_c}         & \ket {i_c}        & \ket {g_t} + \ket {e_t} \\
    \ket 1 & \ket {0_c}         & \ket {g_c}        & \ket {g_t} - \ket {e_t}
  \end{array} \label{compbasis}
\end{equation}

For sake of simplicity we will concentrate on the realization of the
gate assuming that the preparation of the cavity has been already
performed. 

Since all our manipulations are done coherently, it is sufficient to
describe the quantum evolution for basis vectors of the whole system
(ie, our three qubits). The protocol can be decomposed into three
distinct phases: encoding, atomic collision, and decoding. The
encoding and decoding operations only involve resonant interactions
between the cavity mode (control 1) and a single Rb atom (control 2).
Decoding is performed by applying the encoding evolutions in reverse
order. The heart of the evolution is the cavity assisted atomic
collision which realizes the equivalent of the Toffoli gate on the
encoded quantum information. More precisely, it realizes a control
phase gate between the encoded information of the two control qubits
and the target qubit. Thus, the first atom (control 2) is sent at a
relatively low speed into the cavity. After the encoding, it collides
with the faster moving second atom (target). When the second atom has
left the cavity, the first one interacts with the cavity mode to
accomplish the decoding step.

The encoding starts when the atom identified as Control~2 ($A_c$) is
sent into the cavity. First, it interacts resonantly with the cavity
field and undergoes a $\pi$-Rabi rotation. In terms of the basis
vectors given in Eq.~(\ref{compbasis}), only one of them is affected
by the evolution:
\begin{equation} 
\ket {1_c}\ket{g_c} \rightarrow -\ket{0_c}\ket{e_c}.\label{preparation}
\end{equation}
At this point, for a generic input state of the Toffoli gate, all
three levels of the control atom can be populated: the quantum
information which was initially stored into two separated qubits is
now spread over one qubit (the cavity field) and one qutrit (the Rb
atom). The state obtained in Eq.\@~(\ref{preparation}) almost
corresponds to the needed preparation of the cavity and control atom
before the atomic collision: a microwave pulse tuned to the $\ket
{g_c} \leftrightarrow \ket {i_c}$ transition, and corresponding to a
basis rotation exchanging $\ket{i_c} $ and $\ket{g_c}$, completes this
first step of the protocol.

The main part of the protocol can now be achieved: the cavity assisted
atomic collision. Thus, the cavity is set far from resonance with the
$\ket e \leftrightarrow \ket g$ transition, such that the interaction
Hamiltonian is given by Eq.\@~(\ref{Heff}). The target atom ($A_t$)
then enters the cavity and interacts with the field and the control
atom. The evolution of the basis states is easily computed by
diagonalizing the effective Hamiltonian. After an interaction time
$t_{\mathrm {col}} = \pi/\lambda$, all states remain unaffected except
the following ones:
\begin{equation}
\begin{array}{lcl}
\ket{0_c}\ket{i_c}(\ket{g_t} + \ket{e_t}) & \rightarrow & \ket{0_c}\ket{i_c}(\ket{g_t} - \ket{e_t}) \\
\ket{0_c}\ket{i_c}(\ket{g_t} - \ket{e_t}) & \rightarrow & \ket{0_c}\ket{i_c}(\ket{g_t} + \ket{e_t}).\label{collision}
\end{array}
\end{equation}

All operations done before the atomic collision can be considered as
the preparation of the quantum systems such that they undergo the
proper overall evolution described by the Hamiltonian of
Eq.\@~(\ref{collision}). Hence, to complete the whole protocol, we
only need to undo all the preparatory steps: after $A_t$ left the
cavity (recall $A_t$ is the fast moving atom), we apply the resonant
pulse on $A_c$ exchanging $\ket{i_c}$ and $\ket{g_c}$, followed by a
$\pi$-Rabi rotation to extricate from $A_c$ the information initially
contained in the cavity.

\begin{figure}[tbp]
\center

\begin{picture}(0,0)%
\includegraphics{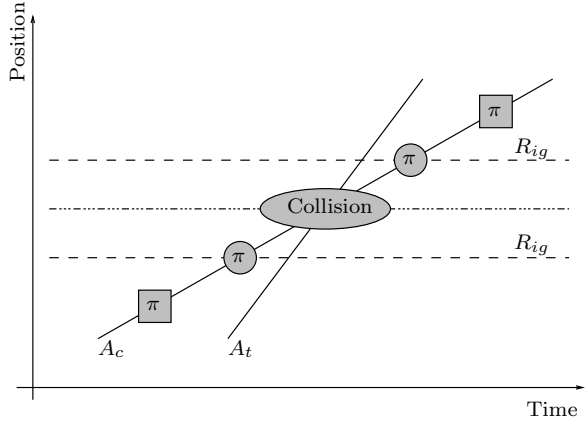}%
\end{picture}%
\setlength{\unitlength}{3947sp}%
\begingroup\makeatletter\ifx\SetFigFont\undefined%
\gdef\SetFigFont#1#2#3#4#5{%
  \reset@font\fontsize{#1}{#2pt}%
  \fontfamily{#3}\fontseries{#4}\fontshape{#5}%
  \selectfont}%
\fi\endgroup%
\begin{picture}(3623,2537)(-13,-1711)
\put(537,-1330){\makebox(0,0)[lb]{\smash{\SetFigFont{8}{9.6}{\familydefault}{\mddefault}{\updefault}{\color[rgb]{0,0,0}$A_c$}%
}}}
\put(3226,-1711){\makebox(0,0)[lb]{\smash{\SetFigFont{8}{9.6}{\familydefault}{\mddefault}{\updefault}{\color[rgb]{0,0,0}Time}%
}}}
\put( 76,389){\rotatebox{90.0}{\makebox(0,0)[lb]{\smash{\SetFigFont{8}{9.6}{\familydefault}{\mddefault}{\updefault}{\color[rgb]{0,0,0}Position}%
}}}}
\put(1353,-1330){\makebox(0,0)[lb]{\smash{\SetFigFont{8}{9.6}{\familydefault}{\mddefault}{\updefault}{\color[rgb]{0,0,0}$A_t$}%
}}}
\put(3151,-61){\makebox(0,0)[lb]{\smash{\SetFigFont{8}{9.6}{\familydefault}{\mddefault}{\updefault}{\color[rgb]{0,0,0}$R_{ig}$}%
}}}
\put(3151,-661){\makebox(0,0)[lb]{\smash{\SetFigFont{8}{9.6}{\familydefault}{\mddefault}{\updefault}{\color[rgb]{0,0,0}$R_{ig}$}%
}}}
\put(842,-1049){\makebox(0,0)[lb]{\smash{\SetFigFont{8}{9.6}{\familydefault}{\mddefault}{\updefault}{\color[rgb]{0,0,0}$\pi$}%
}}}
\put(1378,-743){\makebox(0,0)[lb]{\smash{\SetFigFont{8}{9.6}{\familydefault}{\mddefault}{\updefault}{\color[rgb]{0,0,0}$\pi$}%
}}}
\put(2450,-131){\makebox(0,0)[lb]{\smash{\SetFigFont{8}{9.6}{\familydefault}{\mddefault}{\updefault}{\color[rgb]{0,0,0}$\pi$}%
}}}
\put(2985,175){\makebox(0,0)[lb]{\smash{\SetFigFont{8}{9.6}{\familydefault}{\mddefault}{\updefault}{\color[rgb]{0,0,0}$\pi$}%
}}}
\put(1726,-436){\makebox(0,0)[lb]{\smash{\SetFigFont{8}{9.6}{\familydefault}{\mddefault}{\updefault}{\color[rgb]{0,0,0}Collision}%
}}}
\end{picture}

\caption{\label{scheme} \footnotesize Detailed scheme of the atom-field and
atom-atom interactions in the cavity. Circles are interactions with a
classical microwave field tuned to the $\ket i \leftrightarrow \ket g$
transition. This field is generated in two Ramsey zones denoted
$R_{ig}$. Squares symbolize the resonant interaction with the quantum
field stored into the high-$Q$ cavity. The duration time of each
interaction is set either by controlling the length of the pulse or by
applying a voltage to the mirrors of the cavity to change the
atom-field coupling constant $\Omega$.}
\end{figure}

This completes the overall protocol and leads to:
\begin{equation}
\begin{array}{lcl}
\ket{1_c}\ket{i_c}(\ket{g_t} + \ket{e_t}) & \rightarrow & \ket{1_c}\ket{i_c}(\ket{g_t} + \ket{e_t}) \\
\ket{1_c}\ket{i_c}(\ket{g_t} - \ket{e_t}) & \rightarrow & \ket{1_c}\ket{i_c}(\ket{g_t} - \ket{e_t}) \\
\ket{1_c}\ket{g_c}(\ket{g_t} + \ket{e_t}) & \rightarrow & \ket{1_c}\ket{g_c}(\ket{g_t} + \ket{e_t}) \\
\ket{1_c}\ket{g_c}(\ket{g_t} - \ket{e_t}) & \rightarrow & \ket{1_c}\ket{g_c}(\ket{g_t} - \ket{e_t}) \\
\ket{0_c}\ket{i_c}(\ket{g_t} + \ket{e_t}) & \rightarrow & \ket{0_c}\ket{i_c}(\ket{g_t} + \ket{e_t}) \\
\ket{0_c}\ket{i_c}(\ket{g_t} - \ket{e_t}) & \rightarrow & \ket{0_c}\ket{i_c}(\ket{g_t} - \ket{e_t}) \\
\ket{0_c}\ket{g_c}(\ket{g_t} + \ket{e_t}) & \rightarrow & \ket{0_c}\ket{g_c}(\ket{g_t} - \ket{e_t}) \\
\ket{0_c}\ket{g_c}(\ket{g_t} - \ket{e_t}) & \rightarrow & \ket{0_c}\ket{g_c}(\ket{g_t} + \ket{e_t}),
\end{array}
\end{equation}
which, in turn, exactly corresponds to the Toffoli gate in the
computational basis of Eq.\@~(\ref{compbasis}). The pulses and
interactions sequence are summarized in Fig.\@~\ref{scheme}.

\section{Discussion}

The scheme presented here proposes to use the atomic collision as the
key interaction in the making of the Toffoli gate. The most
fundamental reason that lead to this choice is that using only
resonant interactions would not allow us perform the gate in this very
specific one mode {CQED} setting: it has been shown
\cite{RNOB99a}, that resonant interactions can be used to design CNOT
gates and hence lead to universal quantum computation. However, the
implementations of the Toffoli gate using CNOT gates together with one
qubit gates presented in~\cite{DS94a} would require to address the
qubits separately between each CNOT gate. Our particular setting
forbids such addressing to take place inside the cavity. Hence, to
implement the circuits of \cite{DS94a}, we would need to have more
than one cavity. The counterpart of using the atomic collision is have
to encode and decode the quantum information of the control qubits. In
this proposal, each of these steps is accomplished by a single
interaction involving two quantum systems, and an interaction
involving one quantum system and one classical system. It is easily
shown that this cannot be made simpler, and hence that our proposal is
optimal within the restrictions imposed by the experimental setting.

\begin{figure}[tbp]
\center

\begin{picture}(0,0)%
\includegraphics{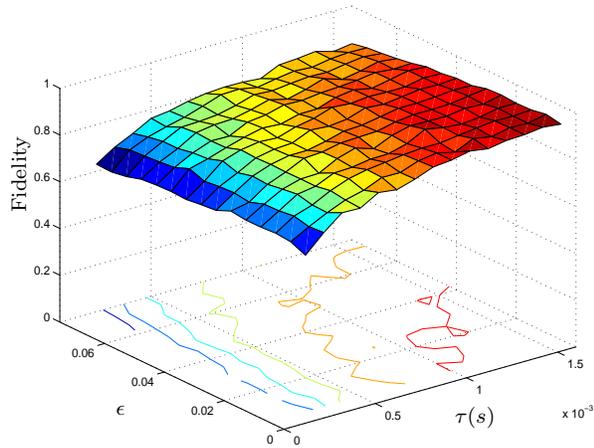}%
\end{picture}%
\setlength{\unitlength}{4144sp}%
\begingroup\makeatletter\ifx\SetFigFont\undefined%
\gdef\SetFigFont#1#2#3#4#5{%
  \reset@font\fontsize{#1}{#2pt}%
  \fontfamily{#3}\fontseries{#4}\fontshape{#5}%
  \selectfont}%
\fi\endgroup%
\begin{picture}(3515,2676)(-139,-1837)
\put(-44,-421){\rotatebox{90.0}{\makebox(0,0)[lb]{\smash{\SetFigFont{8}{9.6}{\familydefault}{\mddefault}{\updefault}{\color[rgb]{0,0,0}Fidelity}%
}}}}
\put(496,-1636){\makebox(0,0)[lb]{\smash{\SetFigFont{8}{9.6}{\familydefault}{\mddefault}{\updefault}{\color[rgb]{0,0,0}$\epsilon$}%
}}}
\put(2521,-1681){\makebox(0,0)[lb]{\smash{\SetFigFont{8}{9.6}{\familydefault}{\mddefault}{\updefault}{\color[rgb]{0,0,0}$\tau(s)$}%
}}}
\end{picture}

\caption{\label{fig:fidelity} \footnotesize Fidelity of the Toffoli gate as a
function of the photon lifetime, $\tau$, and of the uncertainty on the
effective interaction times, $\epsilon$. For current day values, $\tau
= 1\mathrm{ms}$ and $\epsilon = 3\%$ a fidelity of 0.7 is expected.}
\end{figure}

Let us now discuss the practical feasibility of this proposal. The
cavity-field coupling constant is $\Omega / 2\pi =
50\,\mathrm{kHz}$~\cite{RBH01a}. For the atomic collision, we need to
detune the cavity from the $\ket e \leftrightarrow \ket g$ transition
by an amount $\delta \gg \Omega$. This can be done by applying an
external electric field to the mirrors. Here we choose $\delta /2\pi =
4\Omega$, which gives an interaction Hamiltonian well approximated by
Eq\@~(\ref{Heff}) with $\lambda = \Omega^2 / 4\delta$. With these
figures, the interaction time to realize a $\pi$-Rabi rotation is
$2\,10^{-5}\mathrm{s}$ and for the atomic collision $1.25\,
10^{-4}\mathrm{s}$. The time required to perform the interaction with
the classical microwave cavity is negligible. This imposes a velocity
of the atoms of order $50\mathrm{m.s}^{-1}$. This value can be reached
by means of simple atomic beam techniques with transverse laser
cooling. The total interaction time with the cavity field is of order
$0.18 \mathrm{ms}$, still much smaller than the cavity lifetime
($1\mathrm{ms}$). Decoherence effects due to the loss of a photon
should then be relatively small. We present in Fig.~\ref{fig:fidelity}
the result of a numerical simulation to estimate the fidelity of the
gate. Dissipation effects during the atomic collision have been
accounted for using quantum jumps. The fidelity is plotted for various
strength of imperfections in the classical pulses. For a strength of
3\% --- ie, for the achievable precision in the current setting ---
and for cavity lifetime of $1\mathrm{ms}$, the fidelity of the gate is
70\%. Thus, the practical realization of the Toffoli gate through this
protocol is not out of reach and could set an interesting benchmark
for comparing the efficiency of quantum information processing
approaches using {CQED}. Moreover, we can see that even small
improvements on either the precision of the pulses or the cavity
lifetime result in achieving a fidelity of nearly 90\%, thus making
this realization of the gate very attractive for analyzing its
experimental behavior.

We now return to the preparation of the cavity field, before the above
protocol takes place. This can be done in full generality by sending a
Rydberg atom ($A_p$) containing the desired quantum information, and
by transferring its state to the cavity through a $\pi$-Rabi rotation.
The protocol is then started when this ancillary atom leaves the
cavity. The retrieval is accomplished by transferring the state of the
cavity back into an atom $(A_r$) initially in the $\ket g$
state. Thus, the result of the Toffoli gate is contained in the state
of the atoms $A_c$, $A_t$ and $A_r$. This state, and hence the
behavior of the gate, can be easily analyzed by using standard quantum
tomography techniques: this requires only requires single atom
rotations and accumulation of statistics.

\section{Conclusion}

We have presented a realistic scheme for implementing the Toffoli gate
with currently available {CQED} techniques. We have shown that
using a cavity assisted collision instead of only resonant
interactions makes our scheme optimal given the restrictions of the
experimental setting. It thus enlarges the range of possible
applications of CQED to process information for realizing basic
logical operations~\cite{RNOB99a, NROB99a, OBAM01a} as well as more
complicated protocols~\cite{YMBR02a, MOR03a, ZPM03a, SZ02a}. The
estimated achievable fidelity (around 70\% for current imprecision and
decoherence levels) ensures that the behavior of the gate can be
tested experimentally. The corresponding experimental results could be
compared to analogous quantities for other sets of universal gates and
thus provide a deeper insight into the quantum information processing
capability of the {CQED} setting with non resonant
interactions.

\section*{Acknowledgments}
HO was supported in part by the National Security Agency. Laboratoire Kastler Brossel, Universit\'e Pierre et Marie Curie and
ENS, is associated with CNRS (UMR 8552).

\end{document}